\documentstyle[amssymb,twocolumn,pra,aps,psfig]{revtex}

\begin{document}
\title{Full mechanical characterization of a cold damped mirror}
\author{M.\ Pinard\thanks{%
corresponding author: pinard@spectro.jussieu.fr}, P.F.\ Cohadon, T. Briant,
and A.\ Heidmann}
\address{Laboratoire Kastler Brossel\thanks{%
Laboratoire de l'Universit\'{e} Pierre et Marie Curie et de l'Ecole Normale
Sup\'{e}rieure associ\'{e} au Centre National de la Recherche Scientifique},%
\\
Case 74, 4 place Jussieu, F75252\ Paris Cedex 05, France}
\date{July 31, 2000}
\maketitle

\begin{abstract}
We describe an experiment in which we have used a cold damping feedback
mechanism to reduce the thermal noise of a mirror around its mechanical
resonance frequency. The monitoring of the brownian motion of the mirror
allows to apply an additional viscous force without any thermal fluctuations
associated. This scheme has been experimentally implemented with the
radiation pressure of an intensity-modulated laser beam. Large noise
reductions, up to 30 dB, have been obtained. We have also checked the
mechanical response of the cold damped mirror, and monitored its transient
evolution between the cooled regime and the room temperature equilibrium. A
simple theoretical model allows to fully explain the experimental results. A
possible application to the active cooling of the violin modes in a
gravitational-wave interferometer is discussed.\bigskip
\end{abstract}

{\bf PACS :} 05.40.Jc, 04.80.Nn, 42.50.Lc\bigskip

\section{Introduction}

Thermal noise is a major limitation to many very sensitive optical
measurements such as interferometric gravitational wave detection\cite
{VIRGO,LIGO}.\ The suspended mirrors of interferometric detectors have many
degrees of freedom: pendulum modes of the suspension system (with resonance
frequencies below the analysis band), violin modes of the suspension wires
(with some resonances within the detection band), and mirrors' internal
acoustic modes (with resonance frequencies above the band)\cite{Hello}. The
thermal motion of each degree of freedom results from its excitation by a
thermal random force, called the Langevin force. The fluctuation-dissipation
theorem relates the spectrum of the Langevin force to the temperature and to
the mechanical damping of the mode.\ Up to now, only passive methods such as
the increase of mechanical quality factors\cite{Rowan} or cryogenic methods
to lower the temperature\cite{TAMACool} have been considered to lower the
thermal noise and increase the sensitivity.

It has been recently proposed to use a feedback loop to reduce the thermal
noise of a mirror in a high-finesse optical cavity \cite{Tombesi}. A
Fabry-Perot cavity is actually a very sensitive displacement sensor \cite
{EuroPhysics} and a feedback mechanism can be used to control the brownian
motion of a mirror, for example via the radiation pressure of an
intensity-modulated laser beam \cite{PRL}. This approach allows to reduce
the thermal noise both at the mechanical resonance frequencies of the mirror
and at low frequency. At resonance, this cooling corresponds to a cold
damping mechanism, the radiation pressure of the light applying a viscous
force to the mirror without any additional thermal noise.

The purpose of this paper is to present a more detailed theoretical analysis
of the experiment described in\cite{PRL}, focussing on the resonant case,
and to present new experimental results on the mechanical response of the
cold damped mirror and on the transient behavior of the system.

Section \ref{FDT} is dedicated to a brief reminder of the
fluctuation-dissipation theorem and its consequences on the thermal noise
spectrum.\ Section \ref{ColdDamping} presents a detailed theoretical
analysis of the cold damping process used in section \ref{ObsColdDamp} to
lower the effective temperature of the fundamental acoustic mode of a
mirror.\ In section \ref{SecBackground} we take into account the background
thermal noise due to the other acoustic modes of the mirror to explain the
discrepancy between experimental results and the simple monomode theory of
section \ref{ColdDamping}.

We present in section \ref{MecResp} the full mechanical characterization of
the cold damped mirror. We have studied the mechanical response of the
mirror to an external force, when the cooling is applied.\ The results show
that the mechanical susceptibility is changed in agreement with a cold
damping mechanism: the additional radiation pressure corresponds to a
viscous force and there is no additional noise associated with this force,
except the electronic noise of the feedback loop.

Finally, we present the transient evolution of the thermal noise when the
cold damping is switched on or off (section \ref{transient}).\ The results
show that the relaxation time towards the cooled regime can be much shorter
than the relaxation time towards the thermal equilibrium.\ These transient
characteristics may be used to perform a cyclic cooling of the thermal noise
associated with the violin modes of the mirrors in a gravitational-wave
interferometer.

\section{Thermal noise and fluctuation-dissipation theorem}

\label{FDT}We present here some well known results on thermal noise and on
the fluctuation-dissipation theorem, which will give some physical insight
to the efficiency of the feedback scheme used to lower the temperature.

Each degree of freedom is equivalent to a harmonic oscillator, whose thermal
motion is the consequence of the $%
{\frac12}%
k_{B}T$ thermal energy and appears as the mechanical response of the
oscillator to a Langevin force $F_{T}$ describing the coupling to a thermal
bath.\ In the framework of linear response theory \cite{RepLin}, the
resulting motion can be described by its Fourier transform $\delta x\left[
\Omega \right] $ at frequency $\Omega $ which is proportional to $F_{T}\left[
\Omega \right] $, 
\begin{equation}
\delta x\left[ \Omega \right] =\chi \left[ \Omega \right] F_{T}\left[ \Omega %
\right] ,  \label{EqMouvBrownien}
\end{equation}
where $\chi \left[ \Omega \right] $ is the mechanical susceptibility of the
mode, 
\begin{equation}
\chi \left[ \Omega \right] =\frac{1}{M\left( \Omega _{M}^{2}-\Omega
^{2}-i\Omega _{M}^{2}\phi \left( \Omega \right) \right) },  \label{EqChi}
\end{equation}
with a mass $M$, a resonance frequency $\Omega _{M}$, and a loss angle $\phi
\left( \Omega \right) $. In the following, we consider only the case of a
viscous damping, e.g.: 
\begin{equation}
\phi \left( \Omega \right) =\frac{\Gamma \Omega }{\Omega _{M}^{2}},
\label{EqPhiVisc}
\end{equation}
where $\Gamma $ is the mechanical damping of the mode. It is well known that
such a loss angle may not be appropriate to describe some degrees of
freedom, such as internal acoustic modes of the mirrors\cite{SaulsonPRD}: $%
\phi $ is usually considered as independent of frequency over a large
frequency range.\ However, as we are only interested in the spectral
analysis of the motion at frequencies close to a mechanical resonance, for
low loss oscillators ($\phi \ll 1$), both descriptions are equivalent.\ The
former, leading to a simple physical interpretation, will be used throughout
the paper.

The spectrum $S_{F_{T}}$ of the Langevin force is related by the
fluctuation-dissipation theorem\cite{RepLin} to the mechanical
susceptibility, 
\begin{equation}
S_{F_{T}}\left[ \Omega \right] =-\frac{2k_{B}T}{\Omega }%
\mathop{\rm Im}%
\left( \frac{1}{\chi \left[ \Omega \right] }\right) .  \label{SpecLangevin}
\end{equation}
According to Eqs. (\ref{EqMouvBrownien}) to (\ref{SpecLangevin}), the noise
spectrum $S_{x}^{T}\left[ \Omega \right] $ of the displacement $\delta x$ at
temperature $T$ has a lorentzian shape, 
\begin{eqnarray}
S_{x}^{T}\left[ \Omega \right]  &=&\left| \chi \left[ \Omega \right] \right|
^{2}S_{F_{T}}\left[ \Omega \right]   \nonumber \\
&=&\frac{2\Gamma k_{B}T}{M}\frac{1}{\left( \Omega _{M}^{2}-\Omega
^{2}\right) ^{2}+\Gamma ^{2}\Omega ^{2}}.  \label{EqSx}
\end{eqnarray}

Performing the integration of (\ref{EqSx}) over frequency yields the thermal
variance $\Delta x_{T}$, which obeys the equipartition theorem, 
\begin{equation}
{\frac12}%
M\Omega _{M}^{2}\Delta x_{T}^{2}=%
{\frac12}%
k_{B}T.
\end{equation}
If we see the thermal motion as a response to the Langevin force $F_{T}$, it
should be clear that increasing the damping would reduce the thermal
variance.\ However the damping $\Gamma $ cancels from that expression,
because it appears both in the mechanical susceptibility and in the spectrum
of the Langevin force. Decreasing the damping however concentrates the
thermal noise around the mechanical resonance frequency, and therefore
lowers both the low frequency and high frequency thermal noises, which is
the reason for the ongoing works on low loss materials for the substrates of
the mirrors and the suspension wires \cite{Rowan}.

The total noise spectrum, taking into account all the mechanical modes, is
the sum of such lorentzian components.\ Near one of the mechanical
resonances, the thermal noise is mainly ruled by the resonant component,
with a background noise related to all other modes and approximately flat in
frequency.

\section{Cold damping theory}

\label{ColdDamping}The cooling mechanism is based on a measurement of the
thermal noise of the mirror and on a feedback loop which applies a properly
adjusted force on the mirror.\ We will see in section \ref{ObsColdDamp} a
practical implementation of such a mechanism.\ Let us only assume here that
we are able to measure the position $\delta x\left[ \Omega \right] $ of the
mirror, and to apply a viscous feedback force $F_{fb}$, proportional to the
speed $-i\Omega \delta x\left[ \Omega \right] $ of the mirror, 
\begin{equation}
F_{fb}\left[ \Omega \right] =-M\Gamma g\times \left( -i\Omega \delta x\left[
\Omega \right] \right) ,  \label{EqViscForce}
\end{equation}
where $g$ is a dimensionless parameter related to the gain of the feedback
loop. In this section we assume that the mirror can be described as a single
harmonic oscillator. The resulting motion is given by 
\begin{equation}
\delta x\left[ \Omega \right] =\chi \left[ \Omega \right] \left( F_{T}\left[
\Omega \right] +F_{fb}\left[ \Omega \right] \right) ,  \label{EqMouv}
\end{equation}
where $\chi \left[ \Omega \right] $ is the mechanical susceptibility (\ref
{EqChi}) with a loss angle of the form (\ref{EqPhiVisc}), and $F_{T}$ is the
Langevin force whose spectrum is given by the fluctuation-dissipation
theorem (\ref{SpecLangevin}). The motion can be written as 
\begin{equation}
\delta x\left[ \Omega \right] =\chi _{fb}\left[ \Omega \right] F_{T}\left[
\Omega \right] ,  \label{dxChieff}
\end{equation}
where $\chi _{fb}$ is an effective mechanical susceptibility given by 
\begin{equation}
\chi _{fb}\left[ \Omega \right] =\frac{1}{M\left( \Omega _{M}^{2}-\Omega
^{2}-i(1+g)\Gamma \Omega \right) }.  \label{defChieff}
\end{equation}
The additional viscous force (\ref{EqViscForce}) has obviously increased the
damping, but, unlike the case of passive damping, the Langevin force is not
modified and still verifies the fluctuation-dissipation theorem (\ref
{SpecLangevin}): it is only related to the natural damping $\Gamma $ of the
oscillator. We will see shortly that this so called {\it cold damping}
mechanism\cite{ColdDamping1,CD2} allows to reduce the effective temperature
of the system.

Equations (\ref{dxChieff}) and (\ref{defChieff}) show that the mirror now
responds to the Langevin force with the effective mechanical susceptibility $%
\chi _{fb}\left[ \Omega \right] $. As the spectrum of $F_{T}$ is flat
against frequency, the thermal noise spectrum in presence of feedback $%
S_{x}^{fb}\left[ \Omega \right] $ still has a lorentzian shape, of width $%
(1+g)\Gamma $, 
\begin{equation}
S_{x}^{fb}\left[ \Omega \right] =\frac{2\Gamma k_{B}T}{M}\frac{1}{\left(
\Omega _{M}^{2}-\Omega ^{2}\right) ^{2}+(1+g)^{2}\Gamma ^{2}\Omega ^{2}}.
\label{EqSxT}
\end{equation}
As the random force has not been increased, the variance $\Delta x^{2}$ of
the motion is reduced.\ Performing the integration of (\ref{EqSxT}) over the
frequency yields 
\begin{equation}
{\frac12}%
M\Omega _{M}^{2}\Delta x^{2}=\frac{k_{B}T}{2(1+g)}.
\end{equation}
The cold damped oscillator of damping $\Gamma $ and at thermodynamical
temperature $T$ is therefore equivalent to an oscillator with an effective
damping 
\begin{equation}
\Gamma _{fb}=(1+g)\Gamma ,  \label{EqGamma}
\end{equation}
at an effective temperature 
\begin{equation}
T_{fb}=T/(1+g).  \label{EqTfb}
\end{equation}
This can also be deduced from the spectrum of the Langevin force which can
be written as 
\begin{equation}
S_{F_{T}}=2M\Gamma k_{B}T=2M\Gamma _{fb}k_{B}T_{fb},
\end{equation}
where both $\Gamma _{fb}$ and $T_{fb}$ have already been defined. The
increase of the effective damping therefore decreases the effective
temperature of the oscillator. The cold damped mirror is still at
thermodynamical equilibrium but, unlike the case of passive damping, the
temperature is reduced.

We can also define the amplitude noise reduction 
\begin{equation}
R\left[ \Omega \right] =\sqrt{\frac{S_{x}^{T}\left[ \Omega \right] }{%
S_{x}^{fb}\left[ \Omega \right] }}=\left| \frac{\Omega _{M}^{2}-\Omega
^{2}-i\Gamma _{fb}\Omega }{\Omega _{M}^{2}-\Omega ^{2}-i\Gamma \Omega }%
\right| .  \label{EqROmega}
\end{equation}
Off resonance, $R\left[ \Omega \right] $ goes to unity: as the mechanical
susceptibility gets real, the change of its imaginary part does not
contribute anymore to a modification of the spectrum $S_{x}^{fb}\left[
\Omega \right] $. The amplitude noise reduction is maximum at resonance
where it is equal to 
\begin{equation}
R_{T}=R\left[ \Omega _{M}\right] =1+g.  \label{EqRT}
\end{equation}

We have shown in this section that measuring the position of an oscillator
allows the realization of a feedback loop to apply an additional viscous
force to the oscillator without any additional thermal fluctuations.\ Such a
cold damping scheme allows to set the oscillator at thermal equilibrium at a
lower temperature. This is the case only for a single harmonic oscillator.\
We will see in section \ref{SecBackground} that there is no longer a
thermodynamical equilibrium when one takes into account all the acoustic
modes.

\section{Observation of the cold damping}

\label{ObsColdDamp}In this section, we present the experimental
implementation of the feedback scheme presented in the previous section.\
The oscillator is the fundamental acoustic mode of a plano-convex resonator.
We present the experimental results on the cooling at resonance, and the
dependence of the effective damping and temperature with the gain of the
feedback loop.

\subsection{Presentation of the oscillator and motion sensor}

The oscillator used throughout the paper is the fundamental acoustic mode of
a mirror coated on the plane side of a plano-convex resonator made of fused
silica, whose thermal noise observation has been reported in\cite
{EuroPhysics}.\ The mirror has a thickness (at the center) of $1.5\;mm$, a
diameter of $14\;mm$, and a curvature radius of the convex side of $100\;mm$%
. With the value of sound velocity in silica, the value of the thickness
yields a fundamental resonance a bit below $2\;MHz.$

\begin{figure}
\centerline{\psfig{figure=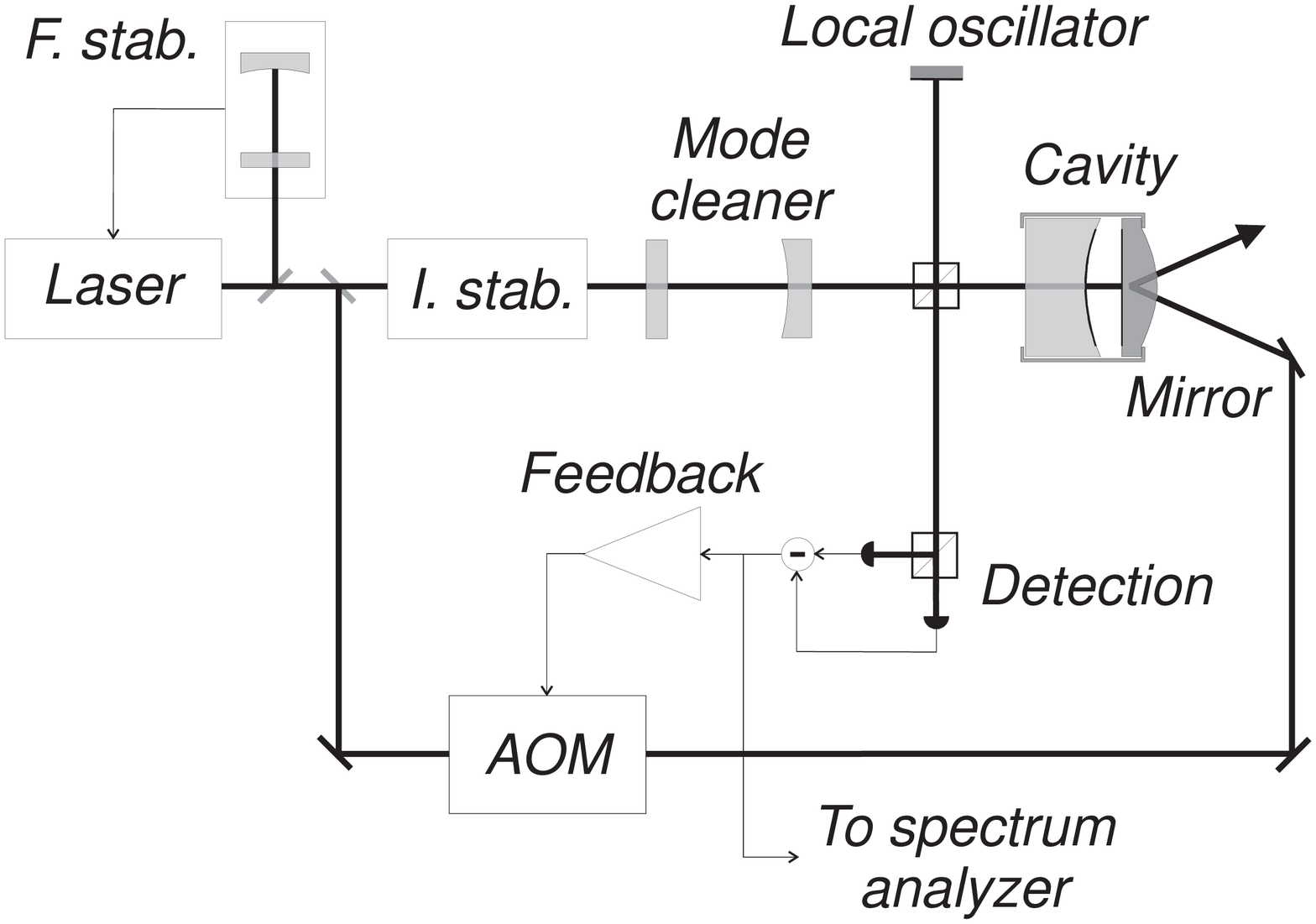,width=7cm}}
\vspace{2mm}
\caption{Experimental setup. The brownian motion of the plano-convex mirror is
measured by a high-finesse cavity. A frequency (F. stab.) and intensity (I. stab.)
stabilized laser beam is sent into the cavity and the phase of the reflected field is
measured by homodyne detection. The radiation pressure of an
intensity-modulated beam is used to excite the motion of the mirror, either to 
characterize its mechanical response or to cool it with a feedback loop.}
\label{Fig_Setup}
\end{figure}

The mirror has been coated at the {\it Institut de Physique Nucl\'{e}aire de
Lyon }and is used as the end mirror of a single-port high-finesse optical
cavity (figure \ref{Fig_Setup}).\ The coupling mirror is a {\it Newport
high-finesse SuperMirror}, held at $1\;mm$ of the rear mirror. The light
entering the cavity is provided by a titane-sapphire laser working at $810~nm
$ and frequency locked to an optical resonance of the high-finesse cavity.\
The light beam is also intensity-stabilized and spatially filtered by a mode
cleaner.

The phase of the field reflected by the cavity is very sensitive to changes
in the cavity length.\ It is monitored by a homodyne detection. The signal
is superimposed to the quantum phase noise of the reflected beam.\ With our
parameters (finesse of the cavity ${\cal F}\simeq 37\;000$, wavelength of
the light $\lambda \simeq 810\;nm$ and incident power $P_{in}=100\;\mu W$),
the quantum limited sensitivity is on the order of $2.7\times 10^{-19}\;m/%
\sqrt{Hz}$\cite{EuroPhysics}. The quantum noise is therefore negligible and
the phase of the reflected field reproduces the brownian motion of the
mirror.

\begin{figure}
\centerline{\psfig{figure=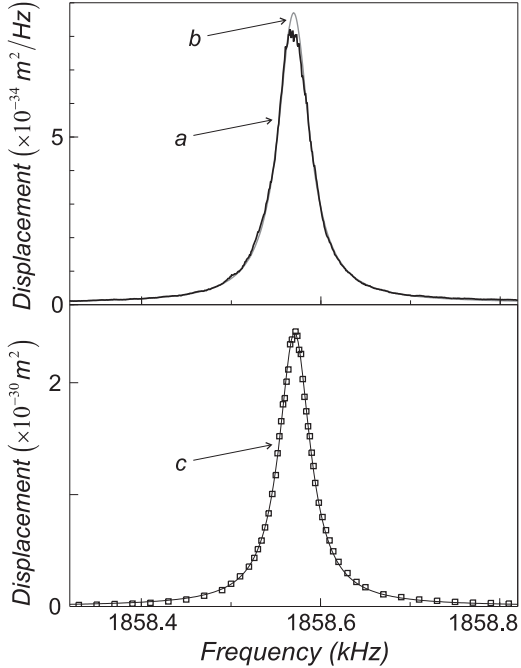,width=6.8cm}}
\vspace{2mm}
\caption{Phase noise spectra of the field reflected by the cavity for a frequency span of
500 Hz around the fundamental resonance frequency of the mirror. Curve {\it a} reflects
the brownian motion of the mirror at room temperature. Curve {\it b} is the theoretical
thermal noise deduced from the mechanical response of the mirror to the optical
excitation (curve  {\it c}; see section \ref{4B}) and from the fluctuation-dissipation
theorem.}
\label{Fig_Thermal}
\end{figure}

Curve (a) of figure \ref{Fig_Thermal} shows the phase noise spectrum of the
reflected field for frequencies around the fundamental resonance frequency
of the mirror.\ The noise is calibrated in displacements of the mirror and
reflects the brownian motion of the mirror which is peaked around the
resonance frequency.\ This experimental result allows to deduce the
mechanical resonance frequency of the fundamental acoustic mode of the
resonator:

\begin{equation}
\Omega _{M}\simeq 2\pi \times 1858.6\;kHz.
\end{equation}
The mechanical susceptibility has a lorentzian shape with a width 
\begin{equation}
\Gamma \simeq 2\pi \times 43\;Hz.
\end{equation}
In other words, the mechanical quality factor $Q=\Omega _{M}/\Gamma $ has a
value of $44\;000$ which seems to be limited by clamping losses. The
effective mass of the mode is deduced from the equipartition theorem, 
\begin{equation}
M\simeq 230\;mg.
\end{equation}

\subsection{Implementation of the feedback loop}

\label{4B}To apply a controlled force on the mirror, we use an auxiliary
laser beam reflected from the back on the mirror (see figure \ref{Fig_Setup}%
).\ This beam is derived from the same laser source as the one used for
motion sensing. The beam is intensity-modulated by an acousto-optic
modulator so that a modulated radiation pressure is applied to the mirror: 
\begin{equation}
F_{rad}\left( t\right) =2\hbar k\;I\left( t\right) ,
\end{equation}
where $k=2\pi /\lambda $ is the wave vector of the laser and $I\left(
t\right) $ is the intensity modulation, expressed as the number of photons
per second. $F_{rad}$ therefore appears as the momentum transferred to the
mirror during the reflection of a single photon, multiplied by the number of
reflections per second.

Driving the acousto-optic modulator with a monochromatic excitation then
allows to characterize the dynamics of the mirror at that frequency. We scan
the excitation frequency for $50$ values spanned around the resonance
frequency in order to characterize the mechanical response of the mirror to
an external force, as this has been done in\cite{EuroPhysics}. The
mechanical response shown in curve (\ref{Fig_Thermal}c) displays a resonance
centered on the same frequency as the thermal noise, and with the same
width.\ Moreover, the observed thermal noise spectrum is in excellent
agreement with the spectrum $S_{x}^{T}\left[ \Omega \right] $ deduced from
this mechanical response and the fluctuation-dissipation theorem (curve \ref
{Fig_Thermal}b) \cite{EuroPhysics}.\ This confirms that the observed thermal
peak is related to the brownian motion of the fundamental acoustic mode of
the plano-convex resonator at room temperature.

Note that the optical excitation is very efficient: with an approximately $%
100\;mW$ completely modulated auxiliary beam, one gets a displacement at
resonance $30\;dB$ larger than the one associated with the thermal noise for
a $1\;Hz$ resolution bandwidth of the spectrum analyzer.

This setup can be used to implement the feedback loop discussed in section 
\ref{ColdDamping}. Indeed, the signal $\delta q_{out}\left[ \Omega \right] $
given by the homodyne detection is proportional to the displacement of the
mirror: 
\begin{equation}
\delta q_{out}\left[ \Omega \right] =a[\Omega ]\delta x\left[ \Omega \right]
,
\end{equation}
where the factor $a\left[ \Omega \right] $ depends on the analysis frequency 
$\Omega $ through the filtering by the cavity.\ In the following, we will
focus on the thermal noise spectrum on a $1\;kHz$ band around the
fundamental resonance frequency.\ This band is small compared to the
bandwidth of the cavity (on the order of $2\;MHz$), so that $a\left[ \Omega %
\right] $ can be considered constant.\ Moreover, this band is small compared
to the mechanical resonance frequency $\Omega _{M}$, so that $\Omega $ can
also be considered constant and\ the derivation necessary to obtain the
force (\ref{EqViscForce}) from the displacement $\delta x\left[ \Omega %
\right] $ can be practically obtained by a dephasing of the homodyne signal.
The feedback loop is therefore simply made of an amplifier with adjustable
gain and phase, in order to drive the acousto-optic modulator with a signal
in quadrature with the displacement $\delta x\left[ \Omega \right] $.

The signal is also filtered by an electronic bandpass filter inserted in the
loop in order to avoid the saturation of the feedback.\ Indeed, the homodyne
signal also comprises resonances associated with higher order modes of the
plano-convex resonator (at higher frequencies) and resonances of the input
mirror (mainly at lower frequencies).\ Moreover, even for the fundamental
acoustic mode considered here, the spectrum of the Langevin force is flat
against frequency and may lead to the saturation of the loop.

As we are only interested in the spectrum around the mechanical resonance
frequency, the filter is centered on this frequency.\ Its bandwidth is
chosen large enough in order to limit its effect on the frequency band of $%
1\;kHz$ around the resonance, and small enough in order to efficiently
filter the low-frequency and high-frequency tails of the resonance.\ The
results presented here have been obtained with a $10\;kHz$ bandwidth, which
fulfills both conditions.\ The influence of the filter can therefore be
neglected in our theoretical model.

\subsection{Noise reduction at resonance}

\begin{figure}
\centerline{\psfig{figure=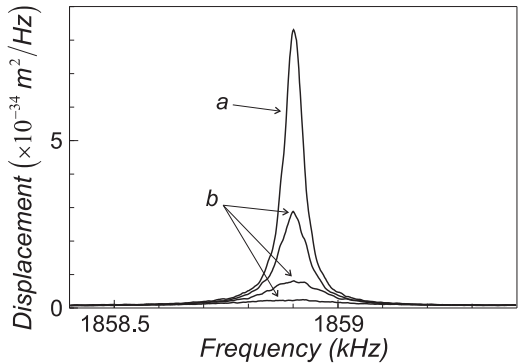,width=6.8cm}}
\vspace{2mm}
\caption{Noise spectra obtained without feedback (curve {\it a}) and with feedback for
increasing values of the gain (curves {\it b}). The temperature, proportional to the area
of the thermal peak, is decreased. The 300 $Hz$ shift in the resonance frequency
relative to figure \ref{Fig_Thermal} is due to a room temperature drift.}
\label{Fig_Cooling}
\end{figure}

Figure \ref{Fig_Cooling} presents the thermal noise spectra obtained without
feedback, and with feedback for increasing values of the gain $g$ of the loop
\cite{PRL}. The thermal noise at the mechanical resonance frequency is
strongly reduced. Meanwhile, as predicted by our theoretical model, the
width of the resonance is strongly increased.\ The effective temperature of
the oscillator is related to the variance $\Delta x^{2}$ and thus to the
area of the curve.\ The decrease of these areas thus corresponds to a
cooling of the mirror.\ We will see in the following that large cooling
factors can be obtained.

\subsection{Dependance of the cooling with the gain}

In order to check our theoretical model, it is instructive to study the
efficiency of the cooling mechanism with respect to the gain $g$ of the
feedback loop. To measure the feedback gain, we detect the intensity of the
auxiliary beam after its reflection on the mirror.\ According to Eq. (\ref
{EqViscForce}), the ratio between the modulation spectrum of the intensity
at frequency $\Omega _{M}$ and the noise spectrum $S_{x}^{fb}\left[ \Omega
_{M}\right] $ of the displacement $\delta x$ leads to the gain $g$, a
multiplicative factor aside. Such a measurement takes into account any
nonlinearity of the gain due to a possible saturation of the acousto-optic
modulator.

\begin{figure}
\centerline{\psfig{figure=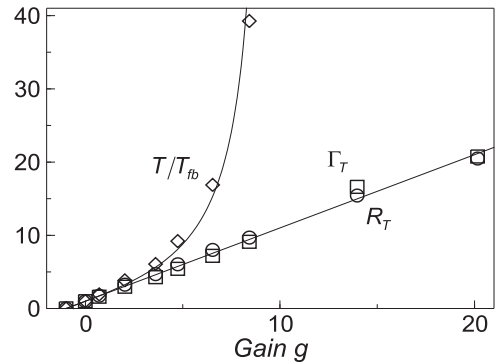,width=6.4cm}}
\vspace{2mm}
\caption{Variation of the relative damping $\Gamma _{T}$ (squares), of the amplitude
noise reduction $R_{T}$ at resonance (circles), and of the cooling factor $T/T_{fb}$
(diamonds), as a function of the feedback gain $g$. Solid lines are theoretical results
(see section \ref{SecBackgroundA}).}
\label{Fig_Gain}
\end{figure}

Figure \ref{Fig_Gain} shows the relative damping $\Gamma _{T}=\Gamma
_{fb}/\Gamma $ and the amplitude noise reduction at resonance $R_{T}$
experimentally observed.\ These parameters are derived from the experimental
spectra by lorentzian fits. As expected from Eqs. (\ref{EqGamma}) and (\ref
{EqRT}), both have a linear dependence with the gain.\ The straight line in
figure \ref{Fig_Gain} is in excellent agreement with experimental data, and
allows to normalize the gain $g$, as this has been done in the figure.
However, the cooling factor $T/T_{fb}$, derived from the area of the
lorentzian fit, does not obey the linear dependence (\ref{EqTfb}). The
following section is dedicated to the explanation of this behavior.

\section{Influence of the background thermal noise}

\label{SecBackground}The purpose of this section is to derive a more
sophisticated theory of our experiment, in order to explain the nonlinear
dependence of the effective temperature with the gain of the loop.\ This
dependence is due to the background thermal noise associated with other
acoustic modes of the mirror or with the input mirror of the cavity.

\subsection{Dependence of the temperature with the gain}

\label{SecBackgroundA}We can take this background noise into account in our
theoretical model.\ Suppose we note $\delta x_{b}\left[ \Omega \right] $ the
associated Fourier component, and $S_{b}\left[ \Omega \right] $ its noise
spectrum.\ As there is no acoustic resonance around the frequency band of
interest, $S_{b}$ is supposed to be frequency independent. The noise $\delta
x_{b}\left[ \Omega \right] $ is uncorrelated to the Langevin force $F_{T}%
\left[ \Omega \right] $ of the fundamental mode, and the overall noise
spectrum without feedback is the sum of the resonant and background
components, 
\begin{equation}
S_{x}^{T}\left[ \Omega \right] =S_{x}^{T(0)}\left[ \Omega \right] +S_{b}.
\label{EqSxBackground}
\end{equation}
where $S_{x}^{T(0)}\left[ \Omega \right] $ is the noise spectrum of the
fundamental mode (Eq. \ref{EqSx}).

Moreover, as the feedback does not involve the input mirror, and as we have
already seen that the efficiency of the loop strongly decreases off
resonance, we can state that the background thermal noise is unaffected by
the loop. We then have 
\begin{equation}
\delta x\left[ \Omega \right] =\chi \left[ \Omega \right] \left( F_{T}\left[
\Omega \right] +F_{fb}\left[ \Omega \right] \right) +\delta x_{b}\left[
\Omega \right] ,
\end{equation}
where $\chi \left[ \Omega \right] $ is the mechanical susceptibility of the
fundamental acoustic mode and the radiation pressure $F_{fb}$ is related to
the overall displacement $\delta x$ by Eq. (\ref{EqViscForce}).\ The
resulting motion is 
\begin{equation}
\delta x\left[ \Omega \right] =\chi _{fb}\left[ \Omega \right] F_{T}\left[
\Omega \right] +\frac{\Omega _{M}^{2}-\Omega ^{2}-i\Gamma \Omega }{\Omega
_{M}^{2}-\Omega ^{2}-i\Gamma _{fb}\Omega }\delta x_{b}\left[ \Omega \right] ,
\end{equation}
where $\chi _{fb}\left[ \Omega \right] $ is the effective susceptibility of
the fundamental mode given by Eq. (\ref{defChieff}). As $\delta x_{b}\left[
\Omega \right] $ and $F_{T}\left[ \Omega \right] $ are uncorrelated, the
noise spectrum takes the simple form: 
\begin{equation}
S_{x}^{fb}\left[ \Omega \right] =S_{x}^{fb(0)}\left[ \Omega \right] +\frac{1%
}{R\left[ \Omega \right] ^{2}}S_{b},  \label{Eq25}
\end{equation}
where $S_{x}^{fb(0)}\left[ \Omega \right] $ is the lorentzian spectrum
obtained when neglecting the background noise (Eq. \ref{EqSxT}), and $R\left[
\Omega \right] $ is the amplitude noise reduction defined in Eq. (\ref
{EqROmega}). Combining Eqs. (\ref{EqROmega}) and (\ref{Eq25}) leads to 
\begin{equation}
S_{x}^{fb}\left[ \Omega \right] =\frac{S_{x}^{T}\left[ \Omega \right] }{R%
\left[ \Omega \right] ^{2}}.
\end{equation}
The feedback loop still decreases the noise spectrum by a factor $R\left[
\Omega \right] ^{2}$. In other words, the effect of the feedback is
unchanged since it is still described by the mechanical susceptibilities $%
\chi $ and $\chi _{fb}$ of the fundamental mode only, but the background
noise in $S_{x}^{T}\left[ \Omega \right] $ leads to a background noise
evenly modified in $S_{x}^{fb}\left[ \Omega \right] $. The noise spectrum (%
\ref{Eq25}) can also be written as 
\begin{equation}
S_{x}^{fb}\left[ \Omega \right] =\left( 1-\varepsilon _{b}g\left( 2+g\right)
\right) S_{x}^{fb(0)}\left[ \Omega \right] +S_{b},  \label{EqSxfb}
\end{equation}
where $\varepsilon _{b}=S_{b}/S_{x}^{T(0)}\left[ \Omega _{M}\right] $ is the
ratio between the background noise spectrum and the resonant component at
frequency $\Omega _{M}$ of the thermal noise $S_{x}^{T}$ without feedback
(Eq.\ \ref{EqSxBackground}). The overall noise spectrum stays lorentzian,
the area of the lorentzian component being smaller when taking $\delta x_{b}%
\left[ \Omega \right] $ into account, with an unchanged background noise.

The width of the resonance is still equal to $\Gamma _{fb}=\left( 1+g\right)
\Gamma $, and the overall amplitude noise reduction is unchanged and equal
to $R\left[ \Omega \right] $ as we have experimentally observed in figure 
\ref{Fig_Gain}.

The effective temperature $T_{fb}$ of the fundamental acoustic mode, related
to the area of the resonant component of the spectrum by the equipartition
theorem, can be computed using Eq.\ (\ref{EqSxfb}): 
\begin{equation}
\frac{T}{T_{fb}}=\frac{1+g}{1-\varepsilon _{b}g(2+g)}.  \label{EqTfbFond}
\end{equation}
The background thermal noise ($\varepsilon _{b}>0$) thus decreases the
effective temperature.

One can experimentally obtain the ratio $\varepsilon _{b}$ using the
lorentzian fit of the thermal noise spectrum observed without feedback
(figure \ref{Fig_Thermal}a). The experimental value is $\varepsilon
_{b}\simeq 1/150.$ The solid curve in figure\ \ref{Fig_Gain} is a fit of the
experimental points using Eq.\ (\ref{EqTfbFond}). The optimum value is $%
\varepsilon _{b}\simeq 1/110$, close to the expected value. The theoretical
curve is in excellent agreement with the experimental points: taking the
background thermal noise into account thus allows to explain the dependence
of the effective temperature with the gain of the feedback loop.

\begin{figure}
\centerline{\psfig{figure=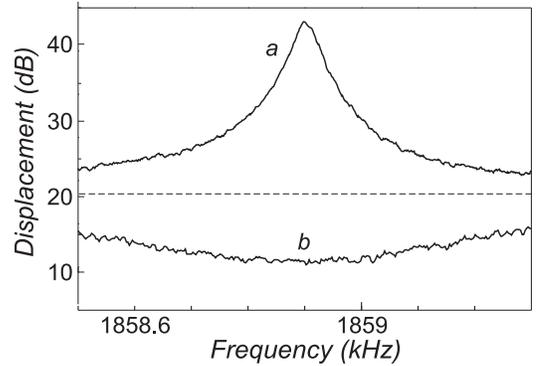,width=6.9cm}}
\vspace{2mm}
\caption{Noise spectra obtained without feedback (curve {\it a}) and with feedback for
a very large gain (curve {\it b}). One gets a dip in the background thermal noise
(dashed line) and the mirror is no longer in a thermodynamical equilibrium.}
\label{Fig_CoolHigh}
\end{figure}

The effective temperature $T_{fb}$ defined previously holds for the resonant
mode, but as the background is unchanged by the feedback loop, one can say
that all the other acoustic modes stay at room temperature $T$. The mirror
is thus no longer in a thermodynamical equilibrium. For sufficient large
gains ($g\gtrsim 10$ in our experiment), the feedback loop can dig a hole in
the background noise, leading to a negative effective temperature for the
resonant mode.\ Figure \ref{Fig_CoolHigh} shows the thermal noise spectrum
experimentally observed for $g\simeq 40$. This figure also shows that very
significant noise reduction can be obtained: the noise power is reduced at
resonance by a factor larger than $1000$.

\subsection{Background noise and thermal equilibrium}

We have discussed in the previous section how the background thermal noise
alters the cooling mechanism. We have used a phenomenological model, where
the background thermal noise is described as an additional noise $\delta
x_{b}\left[ \Omega \right] $ which is unaffected by the feedback loop. If we
neglect the contribution of the coupling mirror to the background noise, it
is possible to derive the same result more formally.

One can show that the motion of the mirror as seen by the intracavity light
can be described by an effective mechanical susceptibility which takes into
account all the acoustic modes and their spatial overlap with the laser
beams \cite{EPJD}: 
\begin{equation}
\chi _{eff}\left[ \Omega \right] =\sum_{n}\chi _{n}\left[ \Omega \right]
\left\langle u_{n},v_{0}\right\rangle ^{2},
\end{equation}
where $\chi _{n}\left[ \Omega \right] $ is the mechanical susceptibility of
the acoustic mode $n$ with spatial deformation $u_{n}\left( {\bf r}\right) $
along the axis of the cavity at every point ${\bf r}$ of the surface, $%
v_{0}\left( {\bf r}\right) $ is the spatial profile of the intensity for the
intracavity field and for the cooling beam (we assume that both beams have
the same profile). $\left\langle f,g\right\rangle $ here stands for the
spatial overlap between two two-dimensional functions, 
\begin{equation}
\left\langle f,g\right\rangle =\int_{S}f({\bf r})g({\bf r})\;d^{2}{\bf r}.
\end{equation}
The thermal motion is then described by an effective Langevin force, which
is a sum of the Langevin forces associated with each acoustic mode.\ One can
show \cite{EPJD} that this Langevin force obeys the fluctuation-dissipation
theorem with respect to $\chi _{eff}$, and the thermal noise spectrum is
therefore given by 
\begin{equation}
S_{x}^{T}\left[ \Omega \right] =\left| \chi _{eff}\left[ \Omega \right]
\right| ^{2}S_{F_{T}}\left[ \Omega \right] =\frac{2k_{B}T}{\Omega }%
\mathop{\rm Im}%
\left( \chi _{eff}\left[ \Omega \right] \right) .  \label{EqStChiEff}
\end{equation}
Around any resonance of the mirror, most of the dynamics is ruled by the
resonant mode, and $\chi _{eff}\left[ \Omega \right] $ mostly displays the
resonant behavior of this mode.

We can now compute the motion of the mirror with the feedback.\ The motion
is as previously 
\begin{equation}
\delta x\left[ \Omega \right] =\chi _{fb}\left[ \Omega \right] F_{T}\left[
\Omega \right] ,  \label{EqdxChiEff}
\end{equation}
where $\chi _{fb}\left[ \Omega \right] $ is the mechanical susceptibility of
the cold damped mirror, 
\begin{equation}
\chi _{fb}\left[ \Omega \right] =\frac{1}{1/\chi _{eff}\left[ \Omega \right]
-iM\Gamma \Omega g}.  \label{Eq33}
\end{equation}

Using Eqs.\ (\ref{EqStChiEff}) and (\ref{EqdxChiEff}), the displacement
noise spectrum takes the form: 
\begin{eqnarray}
S_{x}^{fb}\left[ \Omega \right] &=&\left| \frac{\chi _{fb}\left[ \Omega %
\right] }{\chi _{eff}\left[ \Omega \right] }\right| ^{2}S_{x}^{T}\left[
\Omega \right]  \nonumber \\
&=&\left| \frac{1/\chi _{eff}\left[ \Omega \right] }{1/\chi _{eff}\left[
\Omega \right] -iM\Gamma \Omega g}\right| ^{2}S_{x}^{T}\left[ \Omega \right]
.
\end{eqnarray}
As a first approximation, one can replace the effective susceptibility $\chi
_{eff}\left[ \Omega \right] $ around the fundamental resonance of the mirror
by its resonant component, and find the thermal noise spectrum in presence
of feedback: 
\begin{equation}
S_{x}^{fb}\left[ \Omega \right] =\frac{1}{R\left[ \Omega \right] ^{2}}%
S_{x}^{T}\left[ \Omega \right] ,  \label{Eq35}
\end{equation}
where $R\left[ \Omega \right] $ has already been defined (see Eq.\ \ref
{EqROmega}). As previously, we find that the overall thermal noise spectrum
is reduced by a factor $R\left[ \Omega \right] ^{2}$. The corrections
involved by the off resonance terms in $\chi _{eff}\left[ \Omega \right] $
are negligible for our experimental conditions: the existence of all other
acoustic modes leads to the presence of a background thermal noise in the
noise spectra $S_{x}^{T}\left[ \Omega \right] $ and $S_{x}^{fb}\left[ \Omega %
\right] $, but the dynamics of the cooling mechanism is mainly described by
the resonant component of $\chi _{eff}\left[ \Omega \right] .$

This model also allows to show that the mirror is no longer in
thermodynamical equilibrium in presence of feedback.\ Indeed, this would
require to be able to define an effective temperature $T_{fb}$ obeying the
fluctuation-dissipation theorem at every frequency $\Omega $: 
\begin{equation}
S_{F_{T}}\left[ \Omega \right] =-\frac{2k_{B}T_{fb}}{\Omega }%
\mathop{\rm Im}%
\left( 1/\chi _{fb}\left[ \Omega \right] \right) .  \label{Eq36}
\end{equation}
Using Eqs.\ (\ref{Eq33}) and (\ref{Eq36}), one can show that this requires: 
\begin{equation}
T_{fb}=T/\left[ 1-\frac{M\Gamma \Omega g}{%
\mathop{\rm Im}%
\left( 1/\chi _{eff}\left[ \Omega \right] \right) }\right] .
\end{equation}
This condition can be fulfilled only if $%
\mathop{\rm Im}%
\left( 1/\chi _{eff}\left[ \Omega \right] \right) $ depends linearly of the
frequency $\Omega $.\ This is the case if $\chi _{eff}\left[ \Omega \right] $
describes the mechanical response of a single harmonic viscously-damped
oscillator, but it is no longer the case for a mechanical resonator with
many acoustic modes.

\section{Mechanical response of the cold damped mirror}

\label{MecResp}The experimental results presented in section \ref
{ObsColdDamp} have shown that it is possible to reduce the observed thermal
noise at resonance by a factor $(1+g)$ with the feedback loop. However the
observed quantity is the dephasing of the reflected field, which drives the
feedback loop.\ This error signal goes to zero as the gain of the loop is
increased. In other words, the fact that the thermal noise is reduced does
not prove that the feedback corresponds to a cold damping mechanism.\ To
check if the residual brownian motion of the mirror is given by Eqs.\ (\ref
{dxChieff}) and (\ref{defChieff}) (or Eqs.\ \ref{EqdxChiEff} and \ref{Eq33}
in presence of a background thermal noise), we can determine the cold damped
mechanical susceptibility $\chi _{fb}\left[ \Omega \right] $ by measuring
the mechanical response of the mirror to an external force. This, combined
with the observation of the brownian motion, allows to check that the
Langevin force $F_{T}$ is not altered by the feedback loop, and that the
residual brownian motion is given by Eq. (\ref{dxChieff}).

\subsection{Response to an external force}

The motion of the mirror under the combined influence of the Langevin
thermal force $F_{T}$, the cold damping force $F_{fb}$, and the external
force $F_{ext}$ used to measure its mechanical susceptibility is 
\begin{eqnarray}
\delta x\left[ \Omega \right] &=&\chi _{eff}\left[ \Omega \right] \left(
F_{T}\left[ \Omega \right] +iM\Gamma \Omega g\;\delta x\left[ \Omega \right]
+F_{ext}\left[ \Omega \right] \right)  \nonumber \\
&=&\chi _{fb}\left[ \Omega \right] \left( F_{T}\left[ \Omega \right] +F_{ext}%
\left[ \Omega \right] \right) ,
\end{eqnarray}
where the effective mechanical susceptibility $\chi _{fb}\left[ \Omega %
\right] $ has been defined in Eq. (\ref{Eq33}). In presence of feedback, the
mirror responds the same way to the Langevin force and to an external force.
Checking the mechanical response, one must observe a widening and an
amplitude decrease as the gain $g$ of the loop is increased.

\subsection{Experimental results}

\label{MecRespB}The cold damped mechanical susceptibility $\chi _{fb}\left[
\Omega \right] $ is measured by the same technique already used in \cite
{EuroPhysics}. We use a second auxiliary beam, which is intensity-modulated
by an independent acousto-optic modulator controlled by a high-frequency
oscillator.\ The two auxiliary beams are mixed by a polarizing beamsplitter
before their reflexion on the mirror. The intensity of the second beam is $%
50\;mW$, corresponding to a $30\;dB$ displacement at resonance with respect
to the thermal noise level.\ It is lower than the one of the cooling beam ($%
500\;mW$, $50\;dB$) in order not to alter the cooling mechanism.

\begin{figure}
\centerline{\psfig{figure=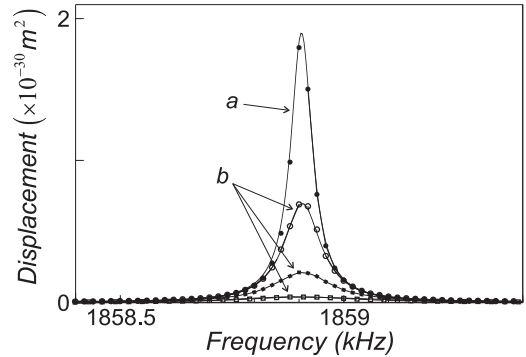,width=6.8cm}}
\vspace{2mm}
\caption{Mechanical response of the mirror to the external force without feedback
(curve {\it a}) and with feedback for increasing values of the gain (curves {\it b}). Solid
lines are lorentzian fits.}
\label{Fig_ModChi}
\end{figure}

For each value of the gain $g$ of the loop, we measure the thermal noise
spectrum, which is used to calibrate $g$.\ We then proceed to a
characterization of the response of the mirror to the external force in
presence of feedback.\ The response is measured for $50$ values of the
modulation frequency of the second auxiliary beam, spanned around the
mechanical resonance.\ Figure\ \ref{Fig_ModChi} represents the modulation
power measured for different values of the gain of the loop.\ Note the
similarity between this figure and figure\ \ref{Fig_Cooling}.

\begin{figure}
\centerline{\psfig{figure=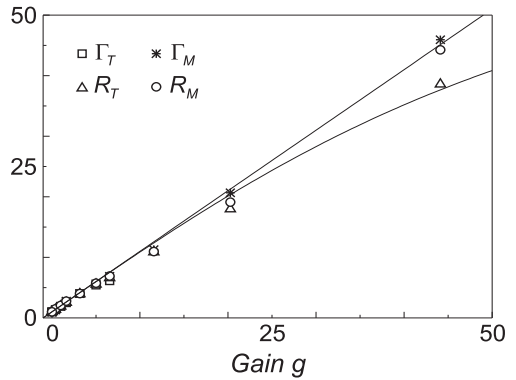,width=6.6cm}}
\vspace{2mm}
\caption{Variation of the relative dampings $\Gamma _{T}$ (squares), $\Gamma _{M}$
(stars), of the amplitude noise reductions at resonance $R_{T}$ (triangles), $R_{M}$
(circles), for the thermal noise and for the mechanical response, as a function of the
feedback gain $g$. Solid lines are theoretical results (see section \ref{MecRespC}).}
\label{Fig_ResChi}
\end{figure}

Each of the curves of figures \ref{Fig_Cooling} and \ref{Fig_ModChi} can be
compared to a lorentzian fit, taking into account the presence of a
background thermal noise.\ One gets the relative width $\Gamma _{T}$ of the
thermal resonance, the relative width $\Gamma _{M}$ of the modulation
resonance, the amplitude thermal noise reduction $R_{T}$ at resonance
(defined in Eq.\ \ref{EqRT}), and the amplitude modulation reduction $R_{M}$
at resonance obtained from the ratio between curves (b) and (a) of figure 
\ref{Fig_ModChi}. Figure \ref{Fig_ResChi} represents the dependence of these
parameters as a function of the gain of the loop ($\Gamma _{T}$ is not
displayed for $g>10$ because the width of the noise spectrum $S_{x}^{fb}$ is
then too large to be accurately measured with a $1\;kHz$ acquisition
bandwidth).

For $g\leq 30$, the experimental points are well aligned on a unity slope
straight line. This confirms that the parameters obey the expected linear
dependence. The susceptibility $\chi _{fb}\left[ \Omega \right] $ perfectly
describes the mechanical behavior of the cold damped mirror.\ Moreover, the
thermal noise spectrum, characterized by $\Gamma _{T}$ and $R_{T}$, still
obeys the fluctuation-dissipation theorem with the mechanical susceptibility 
$\chi _{fb}\left[ \Omega \right] $.\ As stated in our theoretical model, the
Langevin force $F_{T}\left[ \Omega \right] $ is not altered by the cooling
mechanism.

\subsection{Influence of the electronic noise of the loop}

\label{MecRespC}However, for sufficient large gains, the measured thermal
noise reduction $R_{T}$ is lower than expected with our theoretical model.\
This is due to the electronic noise of the feedback loop.\ We can take this
noise into account in our model.\ Suppose we note $\delta x_{e}\left[ \Omega %
\right] $ the electronic noise of the loop. The cold damping force then
takes the form 
\begin{equation}
F_{fb}\left[ \Omega \right] =iM\Gamma \Omega g\left( \delta x\left[ \Omega %
\right] +\delta x_{e}\left[ \Omega \right] \right) ,
\end{equation}
and the resulting motion is 
\begin{equation}
\delta x\left[ \Omega \right] =\chi _{fb}\left[ \Omega \right] \left( F_{T}%
\left[ \Omega \right] +iM\Gamma \Omega g\;\delta x_{e}\left[ \Omega \right]
+F_{ext}\left[ \Omega \right] \right) .
\end{equation}
The response to an external force $F_{ext}\left[ \Omega \right] $ is not
altered by the electronic noise, and it is still described by the effective
susceptibility $\chi _{fb}\left[ \Omega \right] $.\ However, the observed
noise without any external force does depend on the electronic noise $\delta
x_{e}\left[ \Omega \right] $.\ As the electronic noise is uncorrelated to
the Langevin force, and flat in frequency on the frequency band of interest,
the thermal noise spectrum appears as the sum of the thermal noise $%
S_{x}^{fb}\left[ \Omega \right] $ in presence of feedback (see Eq. \ref{Eq35}%
) and the electronic noise spectrum $S_{e}$ injected by the feedback loop, 
\begin{equation}
S_{x}\left[ \Omega \right] =S_{x}^{fb}\left[ \Omega \right] +\left| M\Gamma
\Omega g\chi _{fb}\left[ \Omega \right] \right| ^{2}\;S_{e}.
\label{EqSxEffetBElec}
\end{equation}
The thermal noise spectrum therefore still has a lorentzian shape of width $%
(1+g)\Gamma $, as it is proportional to $\left| \chi _{fb}\left[ \Omega %
\right] \right| ^{2},$ but its amplitude does depend on the electronic
noise.\ The thermal noise at the mechanical resonance frequency is 
\begin{equation}
S_{x}\left[ \Omega _{M}\right] =\frac{S_{x}^{T}\left[ \Omega _{M}\right] }{%
(1+g)^{2}}+\left( \frac{g}{1+g}\right) ^{2}S_{e}.
\end{equation}
The influence of the electronic noise therefore increases with the gain $g$
of the loop. The amplitude noise reduction at resonance is now 
\begin{equation}
R_{T}=\frac{1+g}{\sqrt{1+g^{2}S_{e}/S_{x}^{T}\left[ \Omega _{M}\right] }}.
\end{equation}
The electronic noise can be experimentally measured from the characteristics
of the amplifier used in the loop, and one gets $S_{e}/S_{x}^{T}\left[
\Omega _{M}\right] =2.2\;10^{-4}$.\ The solid curve in figure \ref
{Fig_ResChi} shows the expected dependence of $R_{T}$ with the gain, without
any adjustable parameter.\ It is in excellent agreement with experimental
data.

\section{Transient regime}

\label{transient}We have focussed up to now on the steady-state of the cold
damped mirror. Some questions however arise.\ What is the timescale of the
transition from the natural regime (at thermodynamical temperature $T=300\;K$%
) to the cold damped regime (at effective temperature $T_{fb}$)? And when
the cooling mechanism is switched off? It is well known that the quality
factor $Q$ (and therefore the damping $\Gamma $) can be measured through the
decay time of a mechanical excitation abruptly switched off.\ Is it still
true for the cold damped regime? In this section, we present the
measurements of the timescale needed for the mirror to reach its equilibrium
state, first when the cooling is applied or switched off abruptly, and then
in the case of a mechanical excitation, both for the open-loop and the cold
damped cases.

\subsection{Transition to and from a damped regime}

Suppose we apply sharply the cooling mechanism at $t=0$.\ For $t>0$, the
equation of motion is 
\begin{equation}
M\left[ \ddot{x}+\Gamma \dot{x}+\Omega _{M}^{2}x\right] =F_{T}-M\Gamma g\dot{%
x},  \label{EqEvTemp}
\end{equation}
where $F_{T}(t)$ is the temporal evolution of the Langevin force. The last
term of Eq. (\ref{EqEvTemp}) represents the cold damping force.\ One has to
solve this equation with initial conditions related to the previous
thermodynamical equilibrium at temperature $T$.\ The effect of the cooling
mechanism can be monitored on the variance $\Delta x^{2}(t)$, which is
directly related to the effective temperature by the equipartition theorem.\
We show in Appendix A that the time evolution of the variance is given by 
\begin{equation}
\Delta x^{2}(t)=\frac{\Delta x_{T}^{2}}{1+g}\left( 1+ge^{-\Gamma
_{fb}t}\right) ,  \label{Eqtauon}
\end{equation}
where $\Delta x_{T}^{2}$ is the thermal variance without feedback.\ The
temperature therefore decreases to its new equilibrium value $\Delta
x_{T}^{2}/\left( 1+g\right) $ with a time constant $\Gamma _{fb}^{-1}$. In
other words, the mirror evolves towards its damped regime with a mechanical
response corresponding to its effective susceptibility $\chi _{fb}\left[
\Omega \right] $ in presence of feedback.

In the opposite case, when the cooling is abruptly switched off, the
variance obeys (see Appendix A): 
\begin{equation}
\Delta x^{2}(t)=\Delta x_{T}^{2}\left( 1-\frac{g}{1+g}e^{-\Gamma t}\right) .
\label{EqTauoff}
\end{equation}
The variance goes back to its thermal equilibrium value $\Delta x_{T}^{2}$
with a time constant $\Gamma ^{-1}$ which is now related to the mechanical
damping without feedback.\ The transient evolution to the cooled regime is
then faster by a factor $\Gamma _{fb}/\Gamma =1+g$ than the relaxation
towards the thermal equilibrium.

To observe these transient regimes, the evolution of the variance is
monitored with the spectrum analyzer, centered on the resonance frequency in
the zero-span mode. A fast electronic switch is inserted in the loop in
order to switch alternatively the loop on and off. We have checked that the
transition time of the switch is negligible as compared to the mechanical
time-scales.

The resolution bandwidth of the analyzer must be large enough in order to
reduce the temporal filtering caused by the spectrum analyzer, but not too
large compared to the width of the mechanical resonance, so that the
background noise does not contribute too much to the observed signal. With a
bandwidth of $1\;kHz$, both conditions are fulfilled, the time constant of
the analyzer being on the order of $0.2\;ms$.

\begin{figure}
\centerline{\psfig{figure=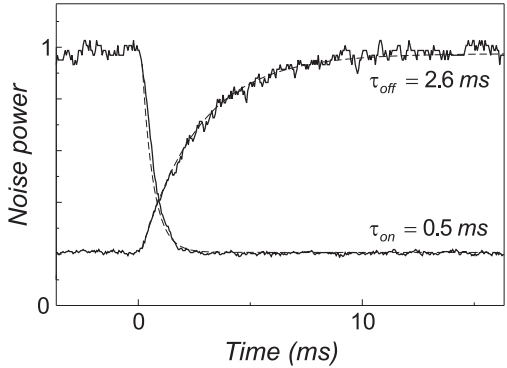,width=6.6cm}}
\vspace{2mm}
\caption{Time evolution of the displacement noise power (in arbitrary units) when the cooling is applied
($\tau _{on}$) and when it is switched off ($\tau _{off}$). Dashed curves are
theoretical fits.}
\label{Fig_TimeCool}
\end{figure}

Figure \ref{Fig_TimeCool} shows the evolution of the variance. When the
cooling is applied, the variance decreases with a time constant $\tau
_{on}\simeq 0.5\;ms$, and when it is switched off, the variance relaxes
towards its equilibrium value with a time constant $\tau _{off}\simeq
2.6\;ms $.\ The experimental data are well fitted by Eqs. (\ref{Eqtauon})
and (\ref{EqTauoff}), with time constants in reasonable agreement with the
expected value: $\Gamma \simeq 2\pi \times 43\;Hz$ leads to a theoretical
value $\tau _{off}\simeq 3.7\;ms$.\ The gain of the loop is here $g=5.2$, so
that the ratio $\tau _{off}/\tau _{on}=5.2$ is also in reasonable agreement
with its expected value of $1+g=6.2$.

\subsection{Decay time for a mechanical excitation}

This paragraph is dedicated to a direct measurement of the damping of the
resonance, through a measurement of the decay time after a mechanical
excitation, both with and without feedback. When a monochromatic force $%
F_{ext}=F_{0}\cos \left( \Omega _{M}t\right) $ is abruptly applied to the
mirror at $t=0$, we show in Appendix B that the displacement spectrum
(without feedback) becomes 
\begin{equation}
S_{x}\left[ \Omega _{M}\right] =%
{\frac14}%
\left( \frac{F_{0}}{M\Omega _{M}}\right) ^{2}\left( 1-e^{-\Gamma t/2}\right)
^{2},  \label{EqSxMecON}
\end{equation}
when the thermal noise is neglected as compared to the effect of the
modulation.\ If the monochromatic excitation is switched off at $t=0$, then
the displacement spectrum is (see Appendix B): 
\begin{equation}
S_{x}\left[ \Omega _{M}\right] =%
{\frac14}%
\left( \frac{F_{0}}{M\Omega _{M}}\right) ^{2}e^{-\Gamma t}.
\label{EqSxMecOFF}
\end{equation}

In presence of feedback, we have already seen that the cold damped mirror is
equivalent to a mirror of intrinsic damping $\Gamma _{fb}$ at temperature $%
T_{fb}.\;$As the thermal noise is neglected, we get similar expressions for $%
S_{x}\left[ \Omega _{M}\right] $ in presence of feedback, replacing $\Gamma $
by $\Gamma _{fb}$.

\begin{figure}
\centerline{\psfig{figure=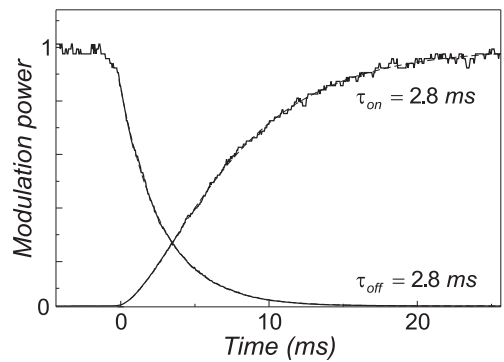,width=6.5cm}}
\vspace{2mm}
\caption{Time evolution of the modulation power of displacement (in arbitrary units) when the external
force is switched on or off, without feedback. Dashed curves are theoretical fits.}
\label{Fig_TimeMod1}
\end{figure}

Figure \ref{Fig_TimeMod1} shows the observed spectra when the excitation is
alternatively switched on and off, without feedback. The modulation power is
measured as previously by the spectrum analyzer in zero-span mode with a
resolution bandwidth of $1~kHz$. The mechanical excitation is applied by the
second auxiliary beam intensity-modulated at frequency $\Omega _{M}$ (see
section \ref{MecRespB}) and switched on and off by the fast electronic
switch. The experimental data are well fitted by Eqs.\ (\ref{EqSxMecON}) and
(\ref{EqSxMecOFF}), with a time constant of $2.8\;ms$, in excellent
agreement with the one found in the previous paragraph.

\begin{figure}
\centerline{\psfig{figure=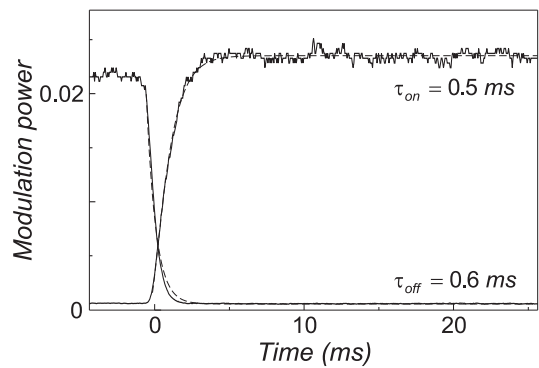,width=7cm}}
\vspace{2mm}
\caption{Time evolution of the modulation power of displacement (same unit
as in figure \ref{Fig_TimeMod1}) when the external
force is switched on or off, with a feedback gain of 5.2. Dashed curves are theoretical
fits.}
\label{Fig_TimeMod2}
\end{figure}

Figure \ref{Fig_TimeMod2} shows the same spectra for a cold damped mirror,
with a gain $g=5.2$.\ The two time constants deduced from the fits with Eqs.
(\ref{EqSxMecON}) and (\ref{EqSxMecOFF}) are $0.5\;ms$ and $0.6\;ms$, both
of the same order of magnitude with the one found in the previous paragraph
in the study of the relaxation towards the damped regime.

We have therefore shown that the relaxation times of the cold damping
mechanism are the ones related to the mechanical response to an external
force, with or without feedback.

\section{Conclusion}

We have demonstrated that it is possible to actively reduce the thermal
noise of a mirror with a feedback loop.\ The cold damping mechanism
described in this paper has allowed us to obtain large noise reductions
around the resonance frequency of the fundamental acoustic mode of the
mirror, up to $30\;dB$. A simple theoretical model, taking into account both
the background thermal noise of the other acoustic modes of the mirror and
the electronic noise of the loop, allows to explain quantitatively the
experimental results. We have also studied the mechanical response of the
cold damped mirror, and the transient regime of the cold damping.

We have in particular studied the mechanical response of the cooled mirror
to an external force, and we have shown that the cooling mechanism indeed
corresponds to a cold damping, where a viscous force is applied on the
mirror without any additional thermal noise but the electronic noise of the
feedback loop.

We have also studied the transient regime of the cold damping mechanism.\ We
have related the relaxation times when the cooling is switched on or off to
the mechanical decay times in response to an external force.\ Our results
show that the cooled regime is much faster to establish than the thermal
equilibrium once the feedback is switched off.

The cold damping might be used to perform a cyclic cooling of thermal peaks
associated with mechanical resonances of the mirror. One can indeed cool a
mechanical resonance during short periods of time, leaving the system
unaffected during much longer periods.\ One then gets a non stationary
thermal noise since it is quickly reduced during the beginning of the cycle
and slowly increases the rest of the time. If the period of the cycle is
properly chosen as compared to the relaxation time of the resonance, the
noise will stay well below its thermal equilibrium value. Furthermore, the
duration of the cooling phase can be very short if the gain of the feedback
is large enough: the system is therefore free most of the time.

Such a cyclic cooling may be used to detect signals corresponding to single
bursts of duration smaller than the period of the cycles. For example, one
can envision to cool the violin modes in a gravitational-wave interferometer
which have sharp resonances within the detection band. Quality factors of
these modes are very large so that the period of the cyclic cooling can be
made much longer that the expected duration of gravitational waves. For a
large feedback gain, the most probable situation is then to detect a
gravitational wave when the cooling is switched off. The gravitational wave
still has the same effect on the interferometer, whereas the thermal noise
is reduced: one then gets a gain in the signal to noise ratio.

\section*{Acknowledgements}

We gratefully acknowledge Yassine {\sc Hadjar} for his work on a previous
stage of this experiment, Jean-Marie {\sc Mackowski} for the optical coating
of the mechanical resonator, and Jean-Michel {\sc Courty} and Serge {\sc %
Reynaud} for stimulating discussions.

\section*{Appendix A}

In this appendix, we study the relaxation of the thermal variance $\Delta
x^{2}\left( t\right) $ between two equilibrium states. In the case where the
cooling is applied at $t=0$, one has to solve equation (\ref{EqEvTemp}) with
initial conditions corresponding to the thermal equilibrium at temperature $T
$. The Laplace transform of (\ref{EqEvTemp}) takes the form 
\begin{equation}
\left[ p^{2}+\Gamma _{fb}p+\Omega _{M}^{2}\right] x\left[ p\right] =\frac{%
F_{T}\left[ p\right] }{M}+\left( \Gamma +p\right) x\left( 0\right) +\dot{x}%
\left( 0\right) ,  \label{eqapp2}
\end{equation}
where $x\left[ p\right] $ and $F_{T}\left[ p\right] $ are respectively the
Laplace transforms of the displacement $x\left( t\right) $ and of the
Langevin force $F_{T}\left( t\right) $. $x\left( 0\right) $ and $\dot{x}%
\left( 0\right) $ are the initial conditions of the oscillator. This leads
to 
\begin{equation}
x\left[ p\right] =\chi _{fb}\left[ p\right] \left( F_{T}\left[ p\right]
+M\left( \Gamma +p\right) x\left( 0\right) +M\dot{x}\left( 0\right) \right) ,
\label{Eqxp}
\end{equation}
where $\chi _{fb}\left[ p\right] $ is the Laplace transform of the effective
susceptibility, 
\begin{equation}
\chi _{fb}\left[ p\right] =\frac{1}{M\left[ \Omega _{M}^{2}+p^{2}+\Gamma
_{fb}p\right] }.  \label{EqChifbp}
\end{equation}
$x\left( t\right) $ can be obtained from Eq. (\ref{Eqxp}) by inverse Laplace
transform.\ It appears as the sum of the response $x_{F}\left( t\right) $ to
the Langevin force (first term in Eq. \ref{Eqxp}) and the transient
evolution $x_{0}\left( t\right) $ from the initial conditions $\left\{
x\left( 0\right) ,\dot{x}\left( 0\right) \right\} $: 
\begin{equation}
x\left( t\right) =x_{F}\left( t\right) +x_{0}\left( t\right) ,
\end{equation}
with 
\begin{mathletters}
\begin{eqnarray}
x_{F}\left( t\right)  &=&\int_{0}^{t}\chi _{fb}\left( t-\tau \right) F\left(
\tau \right) \;d\tau , \\
x_{0}\left( t\right)  &=&M\left( \Gamma x\left( 0\right) +\dot{x}\left(
0\right) \right) \chi _{fb}\left( t\right) +Mx\left( 0\right) \dot{\chi}%
_{fb}\left( t\right) ,  \label{Eqx0}
\end{eqnarray}
where the temporal evolution of $\chi _{fb}\left( t\right) $ is obtained by
inverse Laplace transform of (\ref{EqChifbp}): 
\end{mathletters}
\begin{equation}
\chi _{fb}\left( t\right) =\frac{\sin \left( \Omega _{M}t\right) }{M\Omega
_{M}}\exp \left[ -%
{\frac12}%
\Gamma _{fb}t\right] .  \label{EqCsifb}
\end{equation}

As $F_{T}\left( t>0\right) $ and the initial conditions are uncorrelated,
the variances of $x_{F}$ and $x_{0}$ must be summed, 
\begin{equation}
\Delta x^{2}\left( t\right) =\Delta x_{F}^{2}\left( t\right) +\Delta
x_{0}^{2}\left( t\right) .
\end{equation}
The Langevin force corresponds to a white noise, therefore its correlation
function is 
\begin{equation}
\overline{F_{T}\left( t\right) F_{T}\left( t^{\prime }\right) }=2M\Gamma
k_{B}T\;\delta \left( t-t^{\prime }\right) ,
\end{equation}
and the variance $\Delta x_{F}^{2}\left( t\right) $ is equal to 
\begin{equation}
\Delta x_{F}^{2}\left( t\right) =2M\Gamma k_{B}T\int_{0}^{t}\left[ \chi
_{fb}\left( \tau \right) \right] ^{2}\;d\tau .
\end{equation}
In the limit of a large quality factor of the resonance ($\Gamma \ll \Omega
_{M}$), this leads to 
\begin{equation}
\Delta x_{F}^{2}\left( t\right) =\frac{\Delta x_{T}^{2}}{1+g}\left(
1-e^{-\Gamma _{fb}t}\right) ,
\end{equation}
where $\Delta x_{T}^{2}=k_{B}T/M\Omega _{M}^{2}$ is the variance at room
temperature.

At $t=0$, the mirror is in thermodynamical equilibrium at temperature $T$,
thus we get $\Delta x^{2}\left( 0\right) =\Delta x_{T}^{2}$, $\Delta \dot{x}%
^{2}\left( 0\right) =\Omega _{M}^{2}\Delta x_{T}^{2}$ and $\overline{x\left(
0\right) \dot{x}\left( 0\right) }=0.\;$Combined with the expression of $%
x_{0}\left( t\right) $ (Eq.\ \ref{Eqx0}), this leads to 
\begin{equation}
\Delta x_{0}^{2}\left( t\right) =\Delta x_{T}^{2}e^{-\Gamma _{fb}t}.
\end{equation}
The variance $\Delta x^{2}\left( t\right) $ is finally equal to 
\begin{equation}
\Delta x^{2}\left( t\right) =\frac{\Delta x_{T}^{2}}{1+g}\left(
1+ge^{-\Gamma _{fb}t}\right) .
\end{equation}

One can derive in a similar way the result when the cooling is switched off
at time $t=0$, by taking $g=0$ in Eq. (\ref{EqEvTemp}) and initial
conditions corresponding to the cooled regime.\ One gets 
\begin{equation}
\Delta x^{2}\left( t\right) =\Delta x_{T}^{2}\left( 1-\frac{g}{1+g}%
e^{-\Gamma t}\right) .
\end{equation}

\section*{Appendix B}

This appendix is dedicated to the study of the transient response of the
mirror when an external force is switched on or off. Suppose we apply at $%
t=0 $ an external force $F_{ext}$ to the mirror, without feedback.\ The
displacement $x(t)$ then obeys 
\begin{equation}
M\left[ \ddot{x}+\Gamma \dot{x}+\Omega _{M}^{2}x\right] =F_{ext}+F_{T}.
\end{equation}
If the response to the external force is much larger than the thermal noise,
we can neglect the Langevin force and the initial conditions, which are
related to the brownian motion of the mirror. One then gets 
\begin{equation}
x\left( t\right) =\int_{0}^{t}\chi \left( t-\tau \right) F_{ext}\left( \tau
\right) \;d\tau ,
\end{equation}
where $\chi \left( t-\tau \right) $ is the same as in Eq. (\ref{EqCsifb}),
except for the damping $\Gamma $. For a monochromatic force $F_{ext}\left(
t\right) =F_{0}\cos \left( \Omega _{M}t+\phi \right) $, one gets 
\begin{equation}
x\left( t\right) =\frac{F_{0}}{M\Omega _{M}\Gamma }\sin \left( \Omega
_{M}t+\phi \right) \left[ 1-e^{-\Gamma t/2}\right] .
\end{equation}
As $\Gamma \ll \Omega _{M}$, we have a slow increase of the amplitude of a
monochromatic spectrum.\ In that limit, one can compute the temporal
evolution of the displacement spectrum at resonance, 
\begin{equation}
S_{x}\left[ \Omega _{M}\right] =%
{\frac14}%
\left( \frac{F_{0}}{M\Omega _{M}\Gamma }\right) ^{2}\left[ 1-e^{-\Gamma t/2}%
\right] ^{2}.
\end{equation}

In the case where the monochromatic excitation is switched off at $t=0$, one
has to take into account the initial conditions which correspond to the
forced regime of the oscillator.\ The Laplace transform of the displacement
obeys 
\begin{equation}
\left[ p^{2}+\Gamma p+\Omega _{M}^{2}\right] x\left[ p\right] =\left( \Gamma
+p\right) x\left( 0\right) +\dot{x}\left( 0\right) ,
\end{equation}
where: 
\begin{mathletters}
\begin{eqnarray}
x\left( 0\right) &=&\frac{F_{0}}{M\Omega _{M}\Gamma }\sin \left( \phi
\right) , \\
\dot{x}\left( 0\right) &=&\frac{F_{0}}{M\Gamma }\cos \left( \phi \right) .
\end{eqnarray}
Using the results of Appendix A, one finds 
\end{mathletters}
\begin{eqnarray}
x\left( t\right) &=&M\dot{x}\left( 0\right) \chi \left( t\right) +Mx\left(
0\right) \dot{\chi}\left( t\right)  \nonumber \\
&=&\frac{F_{0}}{M\Omega _{M}\Gamma }\sin \left( \Omega _{M}t+\phi \right)
e^{-\Gamma t/2},
\end{eqnarray}
and the corresponding displacement spectrum is 
\begin{equation}
S_{x}\left[ \Omega _{M}\right] =%
{\frac14}%
\left( \frac{F_{0}}{M\Omega _{M}\Gamma }\right) ^{2}e^{-\Gamma t}.
\end{equation}

\end{document}